\makeatletter
\newcommand{\contraction}[5][1ex]{%
  \mathchoice
    {\contraction@\displaystyle{#2}{#3}{#4}{#5}{#1}}%
    {\contraction@\textstyle{#2}{#3}{#4}{#5}{#1}}%
    {\contraction@\scriptstyle{#2}{#3}{#4}{#5}{#1}}%
    {\contraction@\scriptscriptstyle{#2}{#3}{#4}{#5}{#1}}}%
\newcommand{\contraction@}[6]{%
  \setbox0=\hbox{$#1#2$}%
  \setbox2=\hbox{$#1#3$}%
  \setbox4=\hbox{$#1#4$}%
  \setbox6=\hbox{$#1#5$}%
  \dimen0=\wd2%
  \advance\dimen0 by \wd6%
  \divide\dimen0 by 2%
  \advance\dimen0 by \wd4%
  \vbox{%
    \hbox to 0pt{%
      \kern \wd0%
      \kern 0.5\wd2%
      \contraction@@{\dimen0}{#6}%
      \hss}%
    \vskip 0.5ex
    \vskip\ht2}}

\newcommand{\contraction@@}[3][0.05em]{%
  \hbox{%
    \vrule width #1 height 0pt depth #3%
    \vrule width #2 height 0pt depth #1%
    \vrule width #1 height 0pt depth #3%
    \relax}}
\makeatother

\documentclass[aps,prb,preprint,superscriptaddress,showpacs,showkeys]{revtex4}

\ifx\pdftexversion\undefined
  \usepackage[dvips]{graphics}
\else
  \usepackage[pdftex]{graphics}
\fi

\usepackage{graphicx}
\usepackage{epsfig}
\usepackage{psfrag}
\usepackage{amsmath}

\begin{document}


\title{Single Electron Spin Decoherence by Nuclear Spin Bath: Linked Cluster Expansion Approach.}


\author{S. K. Saikin}
\email[]{ssaikin@physics.ucsd.edu}
\homepage[]{http://www.physics.ucsd.edu/~ssaikin/}
\affiliation{Department of Physics, University of California San
Diego} \affiliation{Department of Physics, Kazan State University,
Russia}
\author{Wang Yao}
\affiliation{Department of Physics, University of California San
Diego}
\author{L. J. Sham}
\affiliation{Department of Physics, University of California San
Diego}


\date{\today}

\begin{abstract}
We develop a theoretical model for transverse dynamics of a single
electron spin interacting with a nuclear spin bath. The approach
allows a simple diagrammatic representation and analytical
expressions of different nuclear spin excitation processes
contributing to electron spin decoherence and dynamical phase
fluctuations. It accounts for nuclear spin dynamics beyond
conventional pair correlation models. As an illustration of the
theory, we evaluated the coherence dynamics of a P donor electron
spin in a Si crystal.
\end{abstract}


\maketitle


\section{Introduction}
An electron-nuclear spin coupling substantially affects electron
spin dynamics in solids. This phenomenon is broadly utilized in
EPR probing of material structures.\cite{Schweiger} However, for
novel technological applications of electron spins
\cite{QC,Spintronics} it produces a major obstacle even at low
temperatures when effects of phonons are suppressed.\cite{Tarucha,
Dutt, Markus, Braun, Barbara}  In particular, it is crucial for a
quantum coherence that is a key issue of a quantum
computation.\cite{QC} Entanglement of an electron spin with a
nuclear spin bath results in an irreversible loss of coherence.
Unlike spin relaxation, \cite{Braun} decoherence process is not
suppressed even in strong magnetic fields. There are several
methods to reduce effects of spin bath, such as isotope
purification, \cite{Tyryshk1} dynamical polarization of nuclear
spins or dynamical decoupling of an electron spin evolution.
\cite{Viola} However, it is not clear if experimentally achievable
values of isotope purification, nuclear polarization or precision
of electron spin control are sufficient to suppress effects of
spin bath beyond the required threshold limit.\cite{Error}
Moreover, for some technologically-important materials these
methods may be inapplicable. For example, isotope purification
cannot be used in GaAs nanostructures, because all stable isotopes
of it have non-zero nuclear spins.  In this connexion theoretical
models of electron-nuclear spin dynamics can provide better
understanding of electron spin decoherence processes and also help
in estimating the effectiveness of coherence control schemes.

In this work we investigate dynamics of a localized electron spin
interacting with a nuclear spin bath at a low temperature regime.
We chose a system with a long spin relaxation time that is of
interest to quantum computing. Until recently, this problem has
been studied using stochastic models of spectral diffusion.
\cite{Anderson, Japan, Salikhov} Their results were verified by
numerous experiments carried out on macroscopic samples. Recently,
emphasis of experimental and theoretical studies has been shifted
to dynamics of single quantum systems, where stochastic models are
inappropriate. Several analytical and numerical approaches based
on quantum dynamics have been developed and used to investigate
different aspects of the problem. \cite{Stamp, Khaetskii,
Merkulov, Saykin, Schliemann, Pershin, Nazarov, Coish, Sousa,
Shenvi, Wang, Dobrovitsky, Witzel, Wang1, Xuedong} Among the
issues addressed in these studies are spin relaxation at low
external fields, \cite{Merkulov, Saykin} effects of nuclear spin
polarization on electron spin dynamics, \cite{Xuedong} dynamical
control for spin decoherence, \cite{Wang1} contributions of high
order nuclear spin correlations into an electron spin echo,
\cite{Witzel} etc. However, many questions are still open. How do
stochastic and dynamical models relate to each other? How does one
characterize short time qubit evolution? \cite{Privman} What are
the reversible and irreversible parts of spin dynamics?
\cite{Pines} What are the relative contributions of different
correlated nuclear spin clusters in electron spin dynamics? How do
nuclear spin correlations grow in time? \cite{Cory}

Here, we demonstrate that diagram techniques developed previously
in studies of Heisenberg ferromagnets \cite{Stinchcombe, Koide,
Callen, VLP, Izyumov, Izyumov_book, YLWang, Baryakhtar,
Baryakhtar1} can be applied to evaluate effects of a nuclear spin
bath on a single electron spin in a high field regime. Our
theoretical approach provides a transparent representation of
different nuclear spin dynamical processes contributing to an
electron spin evolution. It naturally accounts for nuclear spin
excitations beyond pair correlation models. We show that
transverse evolution of an electron spin can be factorized to a
precession in a nuclear Overhauser field and more complex dynamics
due to electron-nuclear spin entanglement.\cite{Wang} A
conventional Hahn echo experiment cancels the phase due to
precession in the nuclear field and also suppresses entanglement
with the nuclear bath. As an illustration, we consider dynamics of
an electron spin localized at phosphorous donor impurity in a Si
crystal. We estimate contributions of 2, 3 and 4 nuclear spin
excitations to electron spin decoherence and discuss effects of
Hahn echo on spin decoherence.

The structure of the paper is as follows. In the next section we
describe a Hamiltonian and discuss assumptions used.
Section~\ref{LCEsec} is devoted to diagrammatic representation of
the decoherence process. In Section~\ref{results} we discuss the
approach and consider an example of P donor electron spin in a Si
crystal. Section~\ref{concl} gives the conclusion. In
Appendix~\ref{A} we provide the spin diagrammatic rules, in
Appendix~\ref{B} we discuss some specific properties of the linked
cluster expansion for spin systems and in Appendix~\ref{C} we give
explicit analytical expressions of some high order nuclear spin
contributions to electron spin dynamics.

\section{Model}
\label{model} We consider the spin of a single electron localized
in a quantum dot or bounded by a donor impurity. We assume that
only one type of nuclear spins with $ I=1/2 $ is presented, though
this assumption can be relaxed within the approach used. In a
strong external magnetic field the Hamiltonian for a single
electron spin coupled by the contact hyperfine interaction
\cite{Slichter} with a system of nuclei can be written as

\begin{equation}
H=\omega_e S^z - \omega_I \sum_i I_i^z + S^z \sum_i{A_i^{\rm{hf}}
I_i^z} + 2S^z \sum_{i \neq j} {B_{ij}^{\rm{hf}} I_i^+ I_j^-}+
\sum_{i \neq j}\{A_{ij}^{\rm{dd}} I_i^z I_j^z + B_{ij}^{\rm{dd}}
I_i^{+} I_j^{-} \} . \label{Hamilt}
\end{equation}

\vskip 0.06 true in \noindent A similar Hamiltonian has been used
in previous studies of the spectral diffusion problem.
\cite{Sousa, Wang} Here, we briefly describe the notations and
assumptions. The first two terms in Eq.~(\ref{Hamilt}) account for
electron and nuclear Zeeman energy level splittings in an external
field, $ \rm{H} $, with the Larmor frequencies $ \omega_e = g^*
\beta \rm{H}/ \hbar $ and $ \omega_I = \gamma \rm{H}$
respectively. The {\it z}-axis is chosen along the magnetic field.
The third and fourth terms originate from the contact hyperfine
interaction. In a strong magnetic field direct electron-nuclear
flip-flop transitions, $~(S^+I_i^-+S^-I_i^+)$, are forbidden by
the energy conservation law. Therefore, beside a small visibility
loss, \cite{Shenvi} this part of the contact interaction
contributes to the effective coupling between nuclear spins only.
\cite{Wang} The coupling coefficients are $ A_i^{\rm{hf}} = ( 8/3
) \pi g_e \beta \gamma |\Psi ( R_i )|^2$ and $B_{ij}^{\rm{hf}} =
A_i^{\rm{hf}} A_j^{\rm{hf}} /2 \omega_e$, where $g_e=2$ and $g^*$
are free and effective electron g-factors, $\beta$ is a Bohr
magneton, $\gamma$ is a nuclear $\gamma$-factor and $\Psi(R_i)$ is
an electron wave function at a position of $i$-th nuclear spin.
The last term in Eq.~(\ref{Hamilt}) represents the secular part of
the nuclear spin dipole-dipole interaction. For the electron
nuclear hyperfine interaction, we consider here only the isotropic
contact part. Effects of the dipolar e-n hyperfine interaction,
resulting in spin echo envelope modulation, \cite{Mims, Fanciulli,
Tyrysk, Itoh} has been discussed elsewhere. \cite{Saikin} The
Hamiltonian~(\ref{Hamilt}) is diagonal in the electron spin.
Therefore, the nuclear spin bath affects transverse electron spin
dynamics only.

The initial state of the electron spin plus the system of nuclear
spins is described by the density matrix $\rho(0)$ at the time
moment, $t=0$, when the electron spin state has been prepared. Two
assumptions are applied to $\rho(0)$. First, we use the standard
approximation of a factorized system and bath,\cite{Kampen} $
\rho(0) = \rho_s^0 \otimes \rho_n^0 $. The electron spin is
initially prepared in the pure state, $(1/\sqrt{2})(|+\rangle +|
-\rangle )$, e.g., by a $\pi/2$ pulse. The second assumption is
that the nuclear spin system is in a pure state that is an
eigenstate of $\sum_i {I_i^z}$ operator. With the latter statement
we neglect by nuclear spin-spin correlations at $t < 0$. Influence
of different initial states of nuclear bath on electron spin
evolution has been discussed in
Refs.~\onlinecite{Schliemann}~and~\onlinecite{Coish} though the
lack of nuclear spin-spin interaction in these papers may affect
their conclusions. It has been argued that the pure spin state
utilized here can be useful for quantum computation purposes
because it does not destroy electron spin coherence at short
timescales.
After
statistical averaging over possible initial configurations the
nuclear spin density matrix is $ \rho_n^0 = \sum {p_n |n \rangle
\langle n | }$, where $p_n$ is a statistical weight of a given
nuclear configuration $|n \rangle =
|\uparrow\uparrow\downarrow\uparrow\downarrow\downarrow...
\rangle$.

The evolution of the up-down component ($+-$) of an electron spin
density matrix can be written as \cite{Wang}

\begin{equation}
\rho_{+-}(t) =\rho_{+-}^0e^{-i \omega_e t} {{\rm
Tr}_n}\{e^{-i(H_{0} + V_{+}) t} \rho_n^0 e^{i(-H_{0}+V_{-}) t}\},
\label{Dmatrix}
\end{equation}

\vskip 0.06 true in \noindent where

\begin{equation}
\begin{array}{l}
 H_{0}=(1/2) \sum_i {A_i^{\rm hf} I_i^z},\\
 V_{\pm}=V_{\rm{dd}} \pm V_{\rm{hf}}, \\
 V_{\rm{dd}} = \sum_{i \neq j}\{A_{ij}^{\rm{dd}} I_i^z I_j^z + B_{ij}^{\rm{dd}}
I_i^{+} I_j^{-} \} \\
 V_{\rm{hf}} = \sum_{i \neq j} {B_{ij}^{\rm{hf}} I_i^+ I_j^-}. \\
 \end{array}
\label{H_ud}
\end{equation} \vskip 0.06 true in

\vskip 0.06 true in \noindent Here, we used the fact that the
Hamiltonian~(\ref{Hamilt}) commutes with $S_z$ and projected it to
the electron spin-up ($+$) and spin-down ($-$) subspaces. The
projected operators are written as two terms, $H_{0}$, which is a
sum of single spin operators, and $V_{\pm}$, which describes
spin-spin interactions. A contribution of the nuclear Zeeman
splitting, the second term in Eq.~(\ref{Hamilt}), is cancelled,
because it commutes with the rest of the Hamiltonian. If the
nuclear Larmor frequency, $\omega_I$, varies from a site to site
due to inhomogeneity of the external field or other factors then
its fluctuating part should be included into $H_{0}$. Below, we
will keep the superscript indexes, (hf) and (dd), in the coupling
constants only if it is not clear from the context what type of
interaction is used.

Using the relations

\begin{equation}
\begin{array}{l}
e^{-i(H_{0} + V_{+}) t}= T \{e^{-i\int_{0}^t
V_{+}(t'-t) dt' } \} e^{-i H_{0} t}, \\
 e^{-i(H_{0} - V_{-}) t}=e^{-i H_{0} t} T \{e^{i\int_0^t V_{-}(t') dt' } \}, \\
\end{array}
\label{Expo}
\end{equation}

\vskip 0.06 true in \noindent where $T\{...\}$ is a time ordering
operator, and our assumption on the initial nuclear spin density
matrix, Eq.~(\ref{Dmatrix}) is transformed to

\begin{equation}
\rho_{+-}(t) =\rho_{+-}^0 e^{-i \omega_e t} \sum_n p_n e^{-i
\omega_n t} \langle n |T \{e^{i\int_0^t V_{-}(t') dt' }\}T \{
e^{-i\int_{0}^t V_{+}(t'-t) dt'} \} |n \rangle, \label{Dmat1}
\end{equation}

\vskip 0.06 true in \noindent where $\omega_n = \langle n | 2 H_0
| n \rangle $ is a contribution of a nuclear Overhauser field to
the electron spin precession frequency and $V_{\pm}(t)$ is
$V_{\pm}$ in an interaction picture defined by $H_0$. For a single
shot measurement of a single electron spin the nuclear bath
contributes to the shift of the electron precession frequency,
$\omega_n$, determined by the initial configuration, and
complicated dynamics due to coupling between nuclear spins that is
described by the bracket $\langle n |...|n \rangle$. The weight
factor, $p_n$, corresponds to a statistical averaging over an
ensemble of electron spins or repeated measurements. For an
ensemble measurement, Eq.~(\ref{Dmat1}) describes a free induction
decay, \cite{Slichter} where the transverse magnetic moment decays
due to an inhomogeneous distribution of spin precession
frequencies and also due to spectral diffusion in the presence of
nuclear spin environment.  Eq.~(\ref{Dmat1}) can be viewed as an
exact formal solution for the electron spin dynamics. In following
sections we will evaluate the term in it between the angular
brackets using the Linked Cluster Expansion (LCE) formalism.
\cite{Mahan, Abricosov}

In Eq.~(\ref{Dmat1}) the product of two exponential operators can
be transformed to a single exponential form, $ T
\{e^{-i\int_{-t}^t \tilde{V}(t') dt' } \} $,  by shifting the time
variable $t'-t \rightarrow t'$ in the second exponent. One can see
that in this case the potential, $\tilde{V}(t')$, is continuous
for the hyperfine-mediated interaction given by Eq.~(\ref{H_ud}),
while for the dipole-dipole nuclear spin-spin interactions it
changes sign at $t'=0$. To avoid operations with discontinuous
potentials we will use a two-exponential form of a bracket in
Eq.~(\ref{Dmat1}) keeping in mind that it can be transformed to a
single exponent.

Schematically, the bracket in Eq.~(\ref{Dmat1}) can be shown as a
two branch propagation, Fig.~\ref{Fig1}. The system propagates on
a first branch with the Hamiltonian $V_{+}(t'-t)$ and after that
on a second branch with $-V_{-}(t')$. This is similar to the
non-equilibrim Green's function approach.\cite{Kadanoff,Keldysh}

The factorization of nuclear spin induced dynamics to the phase
factor acquired in the nuclear Overhauser field and complex
dynamics due to entanglement between bath modes is consistent with
EPR experiments. For example, see Ref.~\onlinecite{Feher}, where
an electron spin resonance line broadening can be resolved to
inhomogeneous and homogeneous parts.

The above procedure also can be applied to evaluate electron spin
dynamics in experiments where the electron spin is flipped by
magnetic pulses. For example, for a Hahn spin echo
 ( $\pi/2 - t - \pi - t - $ echo) the
evolution equation can be written as

\begin{equation}
 \rho_{+-}(2t) =-\rho_{-+}^0 \sum_n p_n \langle n |T\{e^{i\int_0^t V_{+}(t-t') dt' }\}T\{ e^{i\int_0^t V_{-}(t') dt' }\}
 T\{ e^{-i\int_{0}^t V_{+}(t'-t) dt'}\}T\{ e^{-i\int_{0}^t V_{-}(-t') dt'}
\} |n \rangle. \label{bracket}
\end{equation}

\vskip 0.06 true in \noindent We emphasize that the phase factor
due to precession in the external field plus the nuclear field
that appears in Eq.~(\ref{Dmat1}) is cancelled for the echo.
Moreover, the bracket $\langle n |...|n \rangle$ describing
electron-nuclear spin entanglement is different from the one in
Eq.~(\ref{Dmat1}). These features correspond respectively to
elimination of inhomogeneous broadening and suppression of
spectral diffusion in ensemble measurements. \cite{Slichter}

\section{LCE and Decoherence}
\label{LCEsec}
 Using LCE, we can write the expectation value of a time ordered
exponent as \cite{Abricosov, Mahan}

\begin{equation}
\langle n |T \{e^{\int_0^t V (t') dt' } \}|n \rangle = e^{\langle
V_1 \rangle + \langle V_2 \rangle + \langle V_3 \rangle +...},
\label{LCE}
\end{equation}

 \vskip 0.06 true in \noindent where $V (t')$ is an interaction and $ \langle V_k \rangle$ is a
contribution of {\it linked} diagrams \cite{Abricosov} only to the
integral

\begin{equation}
\int_0^t dt_1 \int_0^t dt_2... \int_0^t dt_k \langle n |T \{V
(t_1) V(t_2)...V(t_k) \}|n \rangle . \label{Correl}
\end{equation}

\vskip 0.06 true in \noindent The coefficient, $1/k$, that appears
in LCE, is included in the $ \langle V_k \rangle$ terms. This
expansion provides a convenient exponential form to describe the
dynamical processes. Moreover, each perturbation term $ \langle
V_k \rangle$ in it corresponds to an infinite sum of terms in a
conventional perturbation theory.

The bracket in the expression~(\ref{Correl}) can be evaluated by a
diagrammatic technique. The term $ \langle V_k \rangle $ in
Eq.~(\ref{LCE}) is of the {\it k}-th order in the interaction. It
describes the collective dynamics of a cluster containing up to
{\it k} spins. Because the proof of LCE can be given based on
combinatorics,\cite{Abricosov} the expansion procedure should be
applicable to potentials, discontinuous in time, or to products of
several evolution operators given in
Eqs.~(\ref{Dmat1})-(\ref{bracket}).

 Diagrammatic rules for spins
are not so transparent as for fermions or bosons, because
commutation brackets of spin operators do not yield ${\it
c}$-numbers. Many papers addressed this issue. \cite{Koide,
Callen, VLP, Izyumov, Izyumov_book, YLWang, Baryakhtar,
Baryakhtar1} In our derivations we use the technique described in
Refs.~\onlinecite{Izyumov, Izyumov_book, Baryakhtar1} with
modifications accounting for specifics of the problem. A brief
summary of this technique and the used diagrammatic
representations is given in Appendix~\ref{A}.

The LCE expansion of the scattering matrix~(\ref{LCE}) can be
described by the same series of diagrams as a free energy in
Matsubara formalism given in Ref.~\onlinecite{Izyumov_book}. The
two-exponential representation of the scattering matrix in
Eq.~(\ref{Dmat1}) does not change the structure of the diagram
series, but affects the spin propagators only. In
Figs.~\ref{Fig2}-\ref{Fig4} we show the sets of  diagrams
corresponding to the bracket in Eqs.~(\ref{Dmat1},\ref{bracket})
up to the fourth order in the nuclear spin-spin interaction.

We first discuss the scenario where the nuclear dynamics starts
from a pure initial states. Ensemble results are obtained by
taking statistical average of all possible initial configurations.
Dynamics of each distinct configuration of spins in a cluster, in
general, should be depicted by a different diagram specifying an
initial spin configuration and the time arrow as it is shown in
Fig.~\ref{Fig2}. However, in many cases analytical expressions for
these configurations can be transformed to each other, as is
discussed in Appendix~\ref{A}. To reduce the number of diagrams,
in the figures
 we omit the configuration dependence, assuming that one diagram
 represents all possible configurations. In analytical and numerical
 evaluations we calculate contributions of distinct spin clusters.
 We also drop the time arrow for simplicity.

To evaluate diagrams given in Figs.~\ref{Fig2}-\ref{Fig4} we use a
$2\times2$ matrix (matrix elements indexing branches,
Fig.~\ref{Fig1}) Green's function at an $i$-th site

\begin{equation}
{\bf K }^i(\tau) = e^{i \omega_i \tau} \left(
{\begin{array}{*{20}c}
    \delta_{i \downarrow} \theta (-\tau) -
\delta_{i \uparrow} \theta (\tau) & \delta_{i \downarrow} e^{-i \omega_i t}\\
   -\delta_{i \uparrow} e^{i \omega_i t} &  \delta_{i \downarrow} \theta (-\tau) -
\delta_{i \uparrow} \theta (\tau) \\
\end{array}} \right), \\
\label{Gmat}
\end{equation}

\vskip 0.06 true in \noindent where $ \omega_i=A_i^{\rm hf}/2$ and
$\tau=t_1-t_2$. The total evolution time, $t$, appears in the
off-diagonal elements of ${\bf K}^i(\tau)$. Matrix elements of the
Green's function~(\ref{Gmat}) possess a simple physical meaning.
The propagator starts at time $t_1$ on a branch denoted by a row
index and ends at time $t_2$ on a branch denoted by a column
index. To account for the two-branch propagation the spin-spin
coupling coefficients, $A_{ij}^{\rm dd}, B_{ij}^{\rm dd}$ should
be multiplied by a ${\bf \sigma}_z$ Pauli matrix (${\bf
A}_{ij}^{\rm dd}=A_{ij}^{\rm dd} {\bf \sigma}_z $, ${ \bf
B}_{ij}^{\rm dd}=B_{ij}^{\rm dd}{\bf \sigma}_z$) and $B_{ij}^{\rm
hf}$ should be multiplied by a $2 \times 2$ unity matrix (${\bf
B}_{ij}^{\rm hf}=B_{ij}^{\rm hf} {\bf 1}$). In the analytical
expressions for the diagrams one has to sum over repeating matrix
indexes in addition to integration over the time variables.

The first order linked cluster diagram corresponds to $ \langle n
|{\bf A}_{ij}^{\rm dd} I_i^z I_j^z |n \rangle $ (not shown in
Figs.~\ref{Fig2}-\ref{Fig4}). However, its contribution vanishes
because the sum over the branches is equal to ${\rm Tr} \{ {\bf
A}_{ij}^{\rm dd} \} = 0 $. The first non-zero contribution to the
decoherence process is due to the second order diagrams which
represents a nuclear pair spin flips, Fig.~\ref{Fig2}. The
corresponding analytical expressions are

\begin{equation}
\langle V_2^{\rm hf} \rangle = -\sum_{i=\uparrow, j=\downarrow}
(B_{ij}^{\rm hf})^2 \{\frac{2it}{\omega_{ij}}+\frac{1-e^{2i
\omega_{ij}t}}{\omega_{ij}^2} \},
 \label{V2hf}
\end{equation}

\vskip 0.06 true in \noindent for the hyperfine-mediated
interaction only,  and

\begin{equation}
\langle V_2^{\rm dd} \rangle = -\sum_{i=\uparrow, j=\downarrow}
(B_{ij}^{\rm dd})^2 \{\frac{2it}{\omega_{ij}}+ 4\frac{1-e^{i
\omega_{ij}t}}{\omega_{ij}^2} - \frac{1-e^{2i
\omega_{ij}t}}{\omega_{ij}^2} \},
 \label{V2dd}
 \end{equation}

\vskip 0.06 true in \noindent for the dipole-dipole interaction
only, where we define $\omega_{ij} = \omega_i - \omega_j$. This
result is consistent with Ref.~\onlinecite{Wang}. The real parts
of Eqs.~(\ref{V2hf},\ref{V2dd}) contribute to electron spin
decoherence while the imaginary parts renormalize electron spin
precession frequency and, hence, produce phase fluctuations. The
$\langle V_2 \rangle$ diagram with a cross-term, $B^{\rm hf}B^{\rm
dd}$, contribution of the dipole-dipole and hyperfine-mediated
interactions is zero because the spin propagators over different
branches cancel each other. Therefore, within the second order the
contributions of these two mechanisms to electron spin dynamics
are completely separable.

We emphasize that in comparison with a conventional perturbation
expansion LCE converges faster. Each order in LCE corresponds to
an infinite sum of terms. For example, the second order correction
in LCE is a partial sum shown in Fig.~\ref{Fig5}. It includes a
series of even orders of a perturbation expansion.

 The third order diagrams
in Fig.~\ref{Fig3} can be divided into two groups. First group
includes diagrams~\ref{Fig3}(a), \ref{Fig3}(b) and corresponds to
$\langle V_2 \rangle$ diagram with attached $I_i^z I_j^z$
interaction lines. Diagrams of this type can be accounted for by
renormalization of Green's functions with the standard equation,
Fig.~\ref{Fig6}, where renormalized Green's function (bold line)
is

\begin{equation}
{\bf K }^i(t_1,t_2) = e^{i \omega_i \tau} \left(
{\begin{array}{*{20}c}
    e^{i \Delta \omega_i \tau}(\delta_{i \downarrow} \theta (-\tau) -
\delta_{i \uparrow} \theta (\tau)) & \delta_{i \downarrow} e^{-i \omega_i' t}e^{i \Delta \omega_i \theta}\\
   -\delta_{i \uparrow} e^{i \omega_i' t}e^{-i \Delta \omega_i \theta} &  e^{-i \Delta \omega_i \tau}(\delta_{i \downarrow} \theta (-\tau) -
\delta_{i \uparrow} \theta (\tau)) \\
\end{array}} \right), \\
\label{Gmat_r}
\end{equation}

\vskip 0.06 true in \noindent $\tau=t_1-t_2$, $\theta=t_1+t_2$,
$\omega_{i}'=\omega_{i}+\Delta \omega_i$ and $\Delta
\omega_i=\sum_j {A_{ji} \langle I_j^z \rangle} $. This modifies
the dipole-dipole pair flip-flop term, Eq.~(\ref{V2dd}), as

\begin{equation}
\langle V_2^{\rm ren} \rangle = -\sum_{i=\uparrow, j=\downarrow}
(B_{ij}^{\rm dd})^2
\{\frac{it}{\omega_{11}}+\frac{it}{\omega_{22}}+ \frac{1-e^{i
\omega_{11}t}}{\omega_{11}^2}+ \frac{1-e^{i
\omega_{22}t}}{\omega_{22}^2} + \frac{(1-e^{i
\omega_{11}t})(1-e^{i \omega_{22}t})}{\omega_{11}\omega_{22}} \},
 \label{V2dd_r}
 \end{equation}

\vskip 0.06 true in \noindent where
$\omega_{11}=\omega_{ij}+\Delta \omega_i-\Delta \omega_j$ and
$\omega_{22}=\omega_{ij}-\Delta \omega_i+\Delta \omega_j$. If we
assume that $\Delta \omega_i$ is independent on the site, then
Eq.~(\ref{V2dd_r}) transforms back to Eq.~(\ref{V2dd}). Therefore,
the contribution of these renormalization terms is reduced if the
initial polarization of the nuclear spin bath is homogeneous. A
similar modification of the pair flip-flop term can be due to an
inhomogeneous distribution of nuclear Larmor frequencies,
$\omega_I$ in Eq.~(\ref{Hamilt}). The diagram Fig.~\ref{Fig3}(a)
contributes to renormalization of a pair dynamics by $I^zI^z$
interaction. By direct evaluation one can show that for the
dipole-dipole interaction it is also cancelled .

The second group of diagrams, Fig.~\ref{Fig3}(c), is due to
three-spin flip-flop processes. It corresponds to a ring
propagation of a spin excitation. For the dipole-dipole
interaction only, due to symmetry of interaction terms, the
clockwise propagating excitation cancels the counterclockwise
excitation. It is an analog of the Furry's theorem. \cite{Furry}
An analytical form of the hyperfine-mediated contribution can be
written as

\begin{equation}
\langle V_3^{\rm hf} \rangle = - \sum_{\uparrow \downarrow
\downarrow} B_{ij}B_{jk}B_{ki}
\{\frac{4it}{\omega_{ij}\omega_{ik}}-2 \frac{1-e^{i
\omega_{ik}t}}{\omega_{ik}^2 \omega_{jk}}  +2 \frac{1-e^{i
\omega_{ij}t}}{\omega_{ij}^2 \omega_{jk}}\},
 \label{V3hf}
\end{equation}

\vskip 0.06 true in \noindent for clusters $\{i= \uparrow,j=
\downarrow, k= \downarrow\}$, where permutation of $j$ and $k$
spins is included already. For $\{i=\downarrow, j= \uparrow, k=
\uparrow\}$ clusters, one should change signs at frequencies
$\omega_{i,j,k}$.

In the third order linked diagrams, an effect of cross-terms
including both hyperfine-mediated and dipole-dipole interactions
appears. For example, a ring diagram, Fig.~\ref{Fig3}(c) and
diagram Fig.~\ref{Fig3}(a) with two dipole-dipole interaction
lines and one hyperfine-mediated line have non-zero contributions.
Because most of the third order diagrams give zero contribution to
spin decoherence, the fourth order corrections should be
evaluated.

The set of the fourth order diagrams is given in Fig.~\ref{Fig4}.
The only restriction on vertices in this diagrams is that two
spins $i$ and $j$ coupled by an interaction line should be
different ($i\neq j$). Therefore, the same diagram in
Fig.~\ref{Fig4} can correspond to different number of spins in a
cluster. For example, the ring diagram, Fig.~\ref{Fig4}(f), can
describe excitations of two, three and four spins. We call them
two, three and four-spin ring diagrams respectively.

   In the fourth order in addition to different types of
renormalizations of lower order diagrams, Fig.~\ref{Fig4}(a-e),
and a ring diagram, Fig.~\ref{Fig4}(f) , we have contribution of
locked diagrams, \cite{YLWang} Fig.~\ref{Fig4}(g,h). By locked
diagrams we mean diagrams containing vertices with two incoming
propagators. This group compensates overlapping of spin pair
excitations and also restricts excitation within a spin space
$I^z=\pm 1/2$. The three-spin diagram, Fig.~\ref{Fig4}(g) modifies
a double excitation with one common spin, that appear in expansion
of the exponent of $V_2$ term, and also correct the fourth order
ring diagram with repeating indexes, see Fig.~\ref{Fig7}. The
two-spin diagrams Fig.~\ref{Fig4}(g) and Fig.~\ref{Fig4}(h) plays
a similar role for pairs with two common spins, Fig.~\ref{Fig8}.
In more details we discuss this in Appendix~\ref{B}.

For spin $I=1/2$ two-spin diagram is compensated completely by the
locked diagrams, see Fig.~\ref{Fig8}(b). The tree spin rings
describe dynamics of $i=\uparrow j=\downarrow k=\downarrow$ or
$i=\downarrow j=\uparrow k=\uparrow$ clusters. After correction by
the locked diagram, Fig.~\ref{Fig7}(b), it corresponds to
propagation of a spin excitation as $i \rightarrow j \rightarrow k
\rightarrow j \rightarrow i$. An analytical form of this term
together with locked diagrams, Fig.~\ref{Fig4}(g,h) are given in
Appendix~\ref{C}. A four spin ring diagram correspond to dynamics
of three distinct spin clusters
${\uparrow\downarrow\downarrow\downarrow}$,
$\uparrow\downarrow\uparrow\downarrow$ and
$\uparrow\uparrow\downarrow\downarrow$.

As a result, up to the fourth order in nuclear spin-spin
interactions we write equation for the electron spin coherence as

\begin{equation}
\rho_{+-}(t) =\rho_{+-}^0 e^{-i( \omega_e + \omega_n)t}e^{\langle
V_2 \rangle_n + \langle V_3 \rangle_n + \langle V_4 \rangle_n },
 \label{Dmat_R}
\end{equation}

\vskip 0.06 true in \noindent where the index $n$ denote an
initial configuration of the nuclear bath, the $\langle V_2
\rangle_n$ term (diagram on Fig.~\ref{Fig2}) is given by
Eqs.~(\ref{V2hf},\ref{V2dd}), the $\langle V_3 \rangle_n$ term
(diagram on Fig.~\ref{Fig3}(c)) with expression given by
Eq.~(\ref{V3hf}) and the fourth order contribution of nuclear spin
dynamics schematically shown in Fig.~\ref{Fig9}. Analytical
expressions for some terms of $\langle V_4 \rangle_n$ are in
Appendix~\ref{C}.

The diagram series can be extended to higher orders. At each order
there should be a group of diagrams renormalizing lower orders
with $I^zI^z$ terms, a group of locked diagrams that compensate
overlapping of spin excitations in lower order clusters, and also
a group of ring diagrams.

For systems with a high concentration of nuclear spins,
contributions of different diagrams to a $\langle V_k \rangle$
term can be estimated based on $1/Z$ expansion \cite{Brout}, where
$Z$ is an effective number of interacting spins. For example, if
the last term in Fig.~\ref{Fig9} is O(1), then the first term is
O($1/Z$), second term is O($1/Z$) or O($1/Z^2$) depending on
whether the diagram corresponds to a three or two-spin cluster,
and the third one is O($1/Z^2$). This follows directly from
counting of a number of summands (different spin configurations)
in the analytical expressions, Appendix~\ref{C}. If the effective
number of spins interacting with a given one is large then
$\langle V_4 \rangle$ term can be approximated by the last diagram
in Fig.~\ref{Fig9} only.

\section{Discussion and Example}
\label{results}

The equation for a free evolution of a single electron spin
coupled with a nuclear bath, Eq.~(\ref{Dmat_R}), contains two
terms. First one is a phase factor due to spin precession in the
external field plus the Overhauser field.  The second term is due
to electron-nuclear spin entanglement. In ensemble measurements,
the inhomogeneous distribution of the nuclear Overhauser fields
typically leads to a fast ensemble dephasing time $T_2^*$. This
complicates direct observation of spin decoherence in FID. To
remove the undesired phase factor one can use, for example, a Hahn
spin-echo setup. However, it should be noted that the magnetic
$\pi$ pulse affects the entanglement term also.\cite{Wang} All the
terms evaluated in the previous section can be calculated for the
echo setup straightforwardly. For example, the second order term
with the dipole-dipole interaction, Fig.~\ref{Fig2}, is

\begin{equation}
\langle V_2^{\rm dd} \rangle = -\sum_{i=\uparrow, j=\downarrow}
(B_{ij}^{\rm dd})^2 \{12\frac{1-e^{i \omega_{ij}t}}{\omega_{ij}^2}
+4\frac{1-e^{-i \omega_{ij}t}}{\omega_{ij}^2}- 4\frac{1-e^{2i
\omega_{ij}t}}{\omega_{ij}^2}\}. \label{echo2}
\end{equation}

\vskip 0.06 true in \noindent Moreover, all the diagrams with the
hyperfine-mediated interaction are cancelled for the echo.

As an example we apply the developed technique to a model system,
a phosphorous donor in a Si crystal, that was first studied long
ago. \cite{Feher} Recently, interest in it has been renewed by a
proposal for quantum computation. \cite{Kane} The $^{31}$P is a
shallow donor with the effective radius of the electron wave
function $R_{\rm eff} \sim 25$\AA.\cite{Kohn} Therefore, the
bounded electron covers many host lattice sites. The nuclear spin
bath is represented by a system of randomly distributed $^{29}$Si
isotopes ($I=1/2$). The natural $^{29}$Si isotope concentration is
$c(^{29}{\rm Si}) \approx 4.7\%$. At temperatures $\sim 1$K and
magnetic fields $\sim 0.1-1$T the major mechanism of spin echo
decay in this system is a nuclear spin spectral
diffusion.\cite{Japan,Tyryshk1}

We simulated numerically the processes shown in
Figs.~\ref{Fig2},\ref{Fig9} for a single FID ($\pi/2 - t -$
measurement) and spin echo setup ($\pi/2 -t/2-\pi-t/2-$
measurement). For FID we factorize the shift due to the nuclear
Overhauser field out and concentrate only on the decoherence
induced by electron-nuclear dynamical entanglement. Such
calculation becomes relevant when the inhomogeneous broadening can
be filtered out, e.g., by the method discussed in
Ref.~\onlinecite{Imam}. The contact hyperfine constants for the
system were approximated using the effective mass theory envelop
function.\cite{Kohn} We also assumed that the phosphorus nuclear
spin contributes to the frequency shift only, because of large
difference in $\gamma-$factors of $^{31}$P and $^{29}$Si nuclei.
In simulations of dipole-dipole contributions we generated an
initial nuclear spin configuration in a Si lattice within a sphere
of radius $5R_{\rm eff}$ about the donor. Nuclear spin bath was
assumed un-polarized. Then we selected randomly a spin-up site
with its surrounding within a sphere of a radius $5a$, where
$a=5.43 \AA$ is a Si lattice constant. We calculated contributions
of all possible configurations of a given central spin with its
surrounding. This procedure was repeated for $10^3$ times and
results were normalized to the total number of spin-up within the
whole simulated volume. For the hyperfine-mediated interaction we
averaged over $10^6$ randomly generated spin configurations within
the whole volume. The simulation was done for $100$ different
configurations of nuclear spins to account for initial state
dependence of the decoherence process.  The results were checked
for convergence with changing parameters of the model.

For the electron spin free induction decay the real parts of
different diagrams, Figs.~\ref{Fig2},\ref{Fig9}, averaged over
spatial and spin configurations of $^{29}$Si are shown in
Fig.~\ref{Fig10}. At very short times $t<{\rm max}\{A^{\rm
hf}\}^{-1}$ the second order terms \cite{Wang} are $V_2^{\rm hf}
\sim t^2$, and $V_2^{\rm dd} \sim t^4$. In Fig.~\ref{Fig10}(a) we
show the crossover from the short-time behavior to an intermediate
regime with time dependencies $V_2^{\rm hf} \sim t^{1}, V_2^{\rm
dd} \sim t^{2.3}$. The dispersions in the exponent at intermediate
timescale is $\sim 5 \%$ depending on spatial positions of
$^{29}$Si near the P donor and different initial configurations of
nuclear spins. It also include errors due to a finite simulated
volume. Unlike the electron spin in a quantum dot\cite{Wang} the
short-to-intermediate regime crossover in Si:P is more noticeable
because of the stronger confinement and inhomogeneity of the
electron wave function. For external magnetic fields $H < 0.1-1T$
the hyperfine-mediated term determines short-time spin dynamics,
 see Fig.~\ref{Fig10}(a). However, it can be efficiently suppressed by
increasing the field. Moreover, it cancels completely in the spin
echo. All the dipole-dipole fourth order terms develop on the
timescale of the order of several msec, Fig.~\ref{Fig10}(b).
However, on this timescale the electron spin coherence is
completely destroyed by the $\langle V_2 \rangle$ term. This slow
development of high order spin correlations is consistent with an
experimental measurements.\cite{Cory} At a very short times the
fourth order terms are $ \sim t^6$ (not shown in the figures).
This dependence changes to $V_4^{\rm dd} \sim t^{4.2}$ at longer
timescale. The contribution of the four-spin diagram is about an
order of magnitude larger than other fourth order terms. However,
we abstain from attributing it to 1/Z expansion because there is
no such difference between two and three-spin-fourth order terms.
Probably, this is because the studied system of nuclear spins is
dilute and $Z$ is of order of unity. In Fig.~\ref{Fig11} we show
the total second and fourth order contributions (see inset) in the
exponent for FID and spin echo setups. One can see that a
$\pi$-pulse reduces electron spin decoherence. This effect is an
analog of suppression of a spectral diffusion  considered in
phenomenological models.\cite{Hahn, Slichter}

For the spin echo setup we obtain the time dependence of $V_2^{\rm
dd}$ term comparable to that was calculated in
Ref.~\onlinecite{Witzel}. However, the fourth order terms in our
model do not show the non-monotonic behavior.

We emphasize that in the Si:P system the non-contact hyperfine
coupling between the electron and nuclear spins, not considered
here, produces noticeable effects on electron spin dynamics.
\cite{Fanciulli, Tyrysk, Itoh} To suppress these effects a high
external magnetic field is required. \cite{Saikin} Moreover, in
spin echo measurements on macroscopic samples the dipole-dipole
interaction between electron spins causes an instantaneous
diffusion.\cite{Tyryshk1, Fanciulli1} This effect is beyond the
scope of this paper on a single electron spin. For comparison with
the experimental results given in Ref.~\onlinecite{Tyryshk1}, we
account for the instantaneous diffusion with a phenomenological
exponential decay, $e^{-t/t_{ID}}$. In the simulation we take the
phenomenological relaxation time for instantaneous diffusion to be
$t_{ID}=1.1$ msec, obtained in Ref.~\onlinecite{Tyryshk1}.  The
results are shown in Fig.~\ref{Fig111}. Besides the echo
modulations due to non-contact dipole-dipole interaction observed
in the experiments we still have a moderate discrepancy. Because
the magnetic field used in the experiment was of the order of 0.3
T, the contribution of the hyperfine-mediated terms should be
small on the timescale of the echo decay (0.1-0.5 msec). We
attribute this discrepancy to the effective mass approximation for
the electron wave function. More detailed comparison should be
done after echo modulation and instantaneous diffusion effects are
accounted for by the theory. We leave it for further studies.

\section{Conclusion}
\label{concl} We developed a field theoretic approach to evaluate
dynamics of an electron spin interacting with a nuclear spin bath
in a high field regime. The approach provides a better
understanding of the difference between stochastic models of an
electron spin spectral diffusion and dynamic models of spin
decoherence in the presence of the nuclear spin bath. It also
throws light on the problem of reversibility of spin dynamics. The
approach is based on a conventional diagrammatic technique
utilized in the study of Heisenberg ferromagnets. The scheme
allows for analytical evaluation of different processes
contributing to the electron spin evolution. We show that electron
spin dynamics in a nuclear spin environment can be factorized into
a free precession in the Overhauser field and more complex
dynamics due to an electron-nuclear spin entanglement. The latter
can be evaluated using a linked cluster expansion procedure. The
exact analytical expressions for second order and some high order
processes are given. We show that spin decoherence of a P donor
electron in a single Si crystal is mostly controlled by nuclear
spin pair excitations at sufficiently low temperature and high
magnetic field. Contributions of high order processes are small
and can be neglected on a timescale up to several msec. A magnetic
$\pi$-pulse flipping electron spin slows down the decoherence
process. The simulated results are in fairly good agreement with
experimental measurements of spin echo in macroscopic samples.

\appendix
\section{}
\label{A} We briefly summarize the spin diagrammatic rules. Unlike
Ref.~\onlinecite{Izyumov} where the formalism was used for a
statistically mixed state, our brackets correspond to a pure state
specified by initial conditions. Averaging over a thermal ensemble
in our case would mix dynamical contributions of electron-nuclear
spin entanglement with effects of statistical distribution of
Overhauser fields. For the sake of simplicity here we assume that
$ I=1/2 $. However, the approach can be extended to a more general
case. \cite{YLWang}

The matrix element in Eq.~(\ref{Correl}) can be written as a
product of brackets corresponding to single sites. At each nuclear
spin site $j$ with a given initial state, $|j\rangle$ (=
$|\uparrow \rangle$ or $|\downarrow \rangle$), an expectation
value of time ordered spin operators,

\begin{equation}
 \langle j |T \{I^{\alpha}
(t_1) I^{-}(t_2)...I^{+}(t_k)I^{\beta}(t_{k+1})...
I^{\mu}(t_m)\}|j \rangle \label{Correl1}
\end{equation}

\vskip 0.06 true in \noindent is zero if numbers of $I^{+}$ and
$I^{-}$ operators are not equal. Otherwise, we evaluate it with
the Wick's theorem. \cite{Izyumov} Using the spin commutation
relations, $[I^{-},I^{+}]=-2I^z$ and $[I^{z},I^{+}]=I^{+}$ the
bracket~ (\ref{Correl1}) is transformed to the form where an
operator $I^{+}$ ($I^{-}$ would serve as well) is in the first
position

\begin{equation}
\contraction{\langle j |}{I^{\alpha}}{(t_1)}{I^{+}}
 \langle j | I^{\alpha}(t_1) I^{+}(t)I^{\beta}(t_{2}) |j \rangle  =
\langle j | [I^{\alpha}(t_1),I^{+}(t)]I^{\beta}(t_{2})|j \rangle +
\langle j |I^{+}(t)I^{\alpha}(t_1)I^{\beta}(t_2)|j \rangle
\label{Correl2}
\end{equation}

\vskip 0.06 true in \noindent or in the last position

\begin{equation}
\contraction{\langle j |I^{\alpha}(t_1)}{I^{+}}{(t)}{I^{\beta}}
 \langle j |I^{\alpha}(t_1) I^{+}(t)I^{\beta}(t_{2})|j \rangle  =
-\langle j |I^{\alpha}(t_1),[I^{\beta}(t_{2}),I^{+}(t)]|j \rangle
+ \langle j |I^{\alpha}(t_1)I^{\beta}(t_2) I^{+}(t)|j \rangle .
\label{Correl3}
\end{equation}

\vskip 0.06 true in \noindent Depending on the initial state
operator, $I^+$ is moved to the right if $|j\rangle = |\uparrow
\rangle$ or to the left if $|j\rangle = |\downarrow \rangle$.
After applying

\begin{equation}
\begin{array}{l}
 I^{+}(t)|\uparrow \rangle = 0, \\
\langle \downarrow |I^{+}(t) = 0.
\end{array}
\label{Wick}
\end{equation}

\vskip 0.06 true in \noindent a product of $m$ spin operators is
expanded into a sum of products of $m-1$ operators. This procedure
is repeated until only a product of $I^z$ operators is left. The
latter term is evaluated directly. As a result, the
bracket~(\ref{Correl1}) can be written in terms of all possible
contractors of $I^+$ operator with $I^-$ and $I^z$ using

\begin{equation}
 \begin{array}{l}
[I^{\alpha}(t_1),I^{+}(t)] = e^{i \omega (t-t_1)}
[I^{\alpha},I^{+}]_{t_1}, \label{Pair}
\end{array}
\end{equation}

\vskip 0.06 true in \noindent where the latter commutator is taken
at time $t_1$. Unlike contractions of bosons or fermions a
commutator of $I^+$ and either $I^-$ or $I^z$ is an operator that
can be used in a next pairing. For example, in

\begin{equation}
\contraction{}{I^{+}}{(t)}{I^{z}}
\contraction{I^{+}(t)}{I^{z(t_1)}}{}{I^{-}}
  I^{+}(t)I^{z}(t_1)I^{-}(t_{2})
\label{Exam1}
\end{equation}

\vskip 0.06 true in \noindent the operator $I^{+}(t)$, firstly, is
paired with $I^{z}(t_1)$ with resulting $I^{+}(t_1)$ operator
paired with $I^{-}(t_{2})$. Another specific example of spin
pairing is a locked term. \cite{Izyumov} In

\begin{equation}
 \begin{array}{l}
\contraction{}{I^{+}}{(t)}{I^{-}}
\contraction{I^{+}(t)}{I^{-}(t_1)}{}{I^{+}}
\contraction[2ex]{I^{+}(t)I}{{-}}{(t_1)I^{+}(t')}{I^{-}(t_{2})}
  I^{+}(t) I^{-}(t_1)I^{+}(t')I^{-}(t_{2}),
\label{Exam2}
\end{array}
\end{equation}

\vskip 0.06 true in \noindent the operator $I^{+}(t)$ is paired
with $I^{-}(t_1)$, then $I^{+}(t')$ is paired with the resulting
operator $I^{z}(t_1)$ and finally with $I^{-}(t_{2})$. By the
locked term here we mean a term that contains an operator $I^{-}$
with three contracting lines. The role of such terms in LCE we
discuss in Appendix~\ref{B}.

In diagrams we depict $I^+$ vertices by points, $I^-$ and $I^z$
vertices by open circles and interaction terms by wavy lines. The
Green's function, defined as

\begin{equation}
 K_j(t_1, t_2)=\frac{\langle j |T \{I^{+}
(t_1) I^{-}(t_2)\}|j \rangle}{(-2)\langle j |I^z_j | j \rangle} =
e^{i \omega_j (t_1-t_2)} \{ \delta_{j \downarrow} \theta (t_2 -
t_1) - \delta_{j \uparrow} \theta (t_1 - t_2) \}, \label{Green}
\end{equation}

\vskip 0.06 true in \noindent is shown as a line with an arrow
propagating from $I^+$ to $I^-$. An initial spin state determines
Green's function time evolution. One can see that the Green's
function propagates back in time (arrow points opposite to the
time arrow) if the initial spin state is $\uparrow$, and forward
in time, if the state is $\downarrow$. Although $I^-$ and $I^z$
vertices are depicted by same symbols there is a topological
difference in their appearance in diagrams. $I^z$ vertex can be
either separated from any Green's function or connected to one
incoming and one outgoing Green function. $I^-$ vertex can have
one incoming line or two incoming and one outgoing lines. $I^+$
vertex always has one outgoing Green function line. For
interactions, a wavy line connecting two circles corresponds to
$I^z I^z$ term, while a line connecting a point and a circle
corresponds to $I^+ I^-$ term.

The diagram representation can be easily translated into Green's
functions. For example, the second order flip-flop term,
Fig.~\ref{Fig2} can be written as

\begin{equation}
\langle V_2\rangle = A\sum_{i=\uparrow, j=\downarrow} B_{ij}^2
\int_0^t dt_1 \int_0^t dt_2 K_i(t_1,t_2)K_j(t_2,t_1) (-2)^2
\langle i |I^z_i | i \rangle \langle j |I^z_j | j \rangle.
\label{Exam3}
\end{equation}

\vskip 0.06 true in \noindent The coefficient $A$ in front of the
sum accounts for a number of equivalent diagrams. In the
particular case it is equal one. The coefficient $(-2)^2$ appears
from two contractions of $I^+ I^-$ operators. Analytical
expressions for diagrams are dependent on initial spin states. For
example, pairs $\uparrow \uparrow$ or $\downarrow \downarrow$ give
zero contribution to $\langle V_2 \rangle$ term, see
Fig.~\ref{Fig2}, while contribution of $\uparrow \downarrow$ and
$\downarrow \uparrow$ pairs are equal. We usually omit this
configuration dependence in graphic representation. For example,
three spin diagram, Fig.~\ref{Fig3}(c) corresponds to two possible
spin clusters ${\uparrow \downarrow \downarrow} $ and $\downarrow
\uparrow \uparrow$ (Fig.~\ref{ApA}) that have different analytical
expressions. In general, one can distinguish between
configurations that can be transformed to each other by changing
order in spin counting or by rotation or inversion of the
coordinate system and configurations that are distinct. First type
of configurations are ${\uparrow\downarrow\downarrow}$ and
${\downarrow\uparrow\downarrow}$. If we start counting spins from
the $\uparrow$ site these configurations are the same, and they
have equal analytical expressions. In the case given in
Fig.~\ref{ApA} the two spin configurations are connected by the
inversion operation. An analytical form of the second diagram at
the right hand side can be obtained by changing signs at all
frequencies $\omega$ in an expression for the first diagram.
Distinct configurations are, for example,
${\uparrow\downarrow\uparrow\downarrow}$ and
${\uparrow\uparrow\downarrow\downarrow}$ contributing to the same
forth order ring diagram, Fig.~\ref{Fig3}(f).

\section{}
\label{B} Here we consider in more details a physical origin of
locked terms in LCE. For the sake of simplicity we assume that the
interaction is

\begin{equation}
V(t)=\sum_{ij} B_{ij} I_i^{+}(t) I_j^{-}(t) , \label{inter1}
\end{equation}

\vskip 0.06 true in \noindent and we evaluate

\begin{equation}
 \langle n | T\{e^{-i\int_0^t V(t')dt'} \} |n \rangle = \langle n| 1 +(-i)\int_0^t V(t')dt'+(-i)^2/{2!}\int_0^t \int_0^tT\{ V(t') V(t'')\}dt' dt''+... |n \rangle, \label{inter2}
\end{equation}
\vskip 0.06 true in \noindent where $n$ denotes a spin
configuration. The locked diagrams appear in the fourth order
contribution and correspond to two-spin and three-spin
excitations. A two-spin-fourth-order correction can be written as

\begin{eqnarray}
 3(-i)^4/{4!} \sum_{ij}
B_{ij}^4\int_0^t\int_0^t\int_0^t\int_0^t T\{ \langle i
|I_i^{+}(t_1)I_i^{-}(t_2) I_i^{+}(t_3)I_i^{-}(t_4) |i\rangle \label{inter31}
\\
\nonumber
 \langle j | I_j^{-}(t_1) I_j^{+}(t_2)
 I_j^{-}(t_3) I_j^{+}(t_4)|j \rangle \}  dt_1 dt_2 dt_3 dt_4,
\end{eqnarray}

\vskip 0.06 true in \noindent where we have separated operators
corresponding to different spins. The coefficient 3 in front of
Eq.~(\ref{inter31}) accounts for possible choices of $i$ and $j$.
Eq.~(\ref{inter31}) is nonzero only if $i=\uparrow, j=\downarrow $
(for $i=\downarrow, j=\uparrow $ we just change order in counting
of spins and get the same configuration). The only possible time
ordering in this case is $t_1>t_2>t_3>t_4$ or $t_1>t_4>t_3>t_2$.
The integrand for both sets of time ordering is the same and equal
$e^{i\omega_{ij}(t_1-t_2+t_3-t_4)}$. The same result we can get
with the spin diagram technique and diagram equations given in
Fig.~\ref{Fig8}. Firstly, we expand the time ordered product of
spin operators in Eq.~(\ref{inter31}) in terms of all possible
contractors as discussed in Appendix~\ref{A}. There are two
unlinked terms of the form

\begin{equation}
\contraction{}{I_i^{+}}{(t_1)}{I_i^{-}}
\contraction{I_i^{+}(t_1)I_i^{-}(t_2)}{I_i^{+}}{(t_3)}{I_i^{-}}
\contraction{I_i^{+}(t_1)I_i^{-}(t_2)
I_i^{+}(t_3)I_i^{-}(t_4)}{I_j^{-}}{(t_1)}{I_j^{+}}
\contraction{I_i^{+}(t_1)I_i^{-}(t_2)
I_i^{+}(t_3)I_i^{-}(t_4)I_j^{-}(t_1)I_j^{+}(t_2)
}{I_j^{-}}{(t_3)}{I_j^{+}} I_i^{+}(t_1)I_i^{-}(t_2)
I_i^{+}(t_3)I_i^{-}(t_4)I_j^{-}(t_1)I_j^{+}(t_2) I_j^{-}(t_3)
I_j^{+}(t_4), \label{unlink1}
\end{equation}

\vskip 0.06 true in \noindent two linked terms corresponding to
the ring diagram, given in Fig.~\ref{Fig4}(f)

\begin{equation}
\contraction{}{I_i^{+}}{(t_1)}{I_i^{-}}
\contraction{I_i^{+}(t_1)I_i^{-}(t_2)}{I_i^{+}}{(t_3)}{I_i^{-}}
\contraction{I_i^{+}(t_1)I_i^{-}(t_2)
I_i^{+}(t_3)I_i^{-}(t_4)I_j^{-}(t_1)}{I_j^{-}}{(t_2)}{I_j^{+}}
\contraction[2ex]{I_i^{+}(t_1)I_i^{-}(t_2)
I_i^{+}(t_3)I_i^{-}(t_4)}{I_j^{-}}{(t_1)I_j^{+}(t_2)I_j^{-}(t_3)}{I_j^{+}}
I_i^{+}(t_1)I_i^{-}(t_2)
I_i^{+}(t_3)I_i^{-}(t_4)I_j^{-}(t_1)I_j^{+}(t_2) I_j^{-}(t_3)
I_j^{+}(t_4),\label{ring}
\end{equation}

\vskip 0.06 true in \noindent eight single-site locked terms,
Fig.~\ref{Fig4}(g), of the form

\begin{equation}
\contraction{}{I_i^{+}}{(t_1)}{I_i}
\contraction{I_i^{+}(t_1)I_i^{-}}{(}{t_2)}{I_i^{+}}
\contraction[2ex]{I_i^{+}(t_1)I_i}{^{-}}{(t_2)
I_i^{+}(t_3)}{I_i^{-}} \contraction{I_i^{+}(t_1)I_i^{-}(t_2)
I_i^{+}(t_3)I_i^{-}(t_4)}{I_j^{-}}{(t_1)}{I_j^{+}}
\contraction{I_i^{+}(t_1)I_i^{-}(t_2)
I_i^{+}(t_3)I_i^{-}(t_4)I_j^{-}(t_1)I_j^{+}(t_2)
}{I_j^{-}}{(t_3)}{I_j^{+}} I_i^{+}(t_1)I_i^{-}(t_2)
I_i^{+}(t_3)I_i^{-}(t_4)I_j^{-}(t_1)I_j^{+}(t_2) I_j^{-}(t_3)
I_j^{+}(t_4),\label{1lock}
\end{equation}

\vskip 0.06 true in \noindent and four two-sites locked terms,
Fig.~\ref{Fig4}(h)

\begin{equation}
\contraction{}{I_i^{+}}{(t_1)}{I_i}
\contraction{I_i^{+}(t_1)I_i^{-}}{(}{t_2)}{I_i^{+}}
\contraction[2ex]{I_i^{+}(t_1)I_i}{^{-}}{(t_2)
I_i^{+}(t_3)}{I_i^{-}} \contraction{I_i^{+}(t_1)I_i^{-}(t_2)
I_i^{+}(t_3)I_i^{-}(t_4)I_j^{-}(t_1)}{I_j}{^{+}(t_2)}{I_j}
\contraction{I_i^{+}(t_1)I_i^{-}(t_2)
I_i^{+}(t_3)I_i^{-}(t_4)I_j^{-}(t_1)I_j^{+}(t_2)
I_j^{-}}{(}{t_3)}{I_j^{+}}
\contraction[2ex]{I_i^{+}(t_1)I_i^{-}(t_2)
I_i^{+}(t_3)I_i^{-}(t_4)}{I_j^{-}(t_1)}{I_j^{+}(t_2) I_j}{^{-}}
I_i^{+}(t_1)I_i^{-}(t_2)
I_i^{+}(t_3)I_i^{-}(t_4)I_j^{-}(t_1)I_j^{+}(t_2) I_j^{-}(t_3)
I_j^{+}(t_4).\label{2lock}
\end{equation}

\vskip 0.06 true in \noindent Eq.~(\ref{unlink1}) corresponds to
the series expansion of the second order contribution. However, it
allows some non-physical states, because the contractions in it
specify time ordering $t_1>t_2$ and $t_3>t_4$ only. By direct
evaluation, one can show that the locked terms,
Eqs.~(\ref{1lock},\ref{2lock}), modify it according to the diagram
equation given in Fig.~\ref{Fig8}(a) to get a correct time
ordering $t_1>t_2>t_3>t_4$. It is a correction of overlapping of
spin excitations discussed in Refs.~\onlinecite{Wang,Witzel}. The
locked diagram is the price we pay to obtain LCE for spins. This
shows that the procedure to get an exponential form of a qubit
decoherence is not as transparent as it was suggested in
Ref.~\onlinecite{Witzel}.

\section{}
\label{C} Here we provide explicit analytical expressions for two
and three spin diagrams given in Fig.~\ref{Fig9}. The first term
is denoted as $V_{43}$, the three-spin part of the second term is
$V_{41l}$ and the two-spin part of the second term plus the third
term is $V_{42l}$. We do not include the coefficients 1/2
indicated in Fig.~\ref{Fig9}. The analytical expression for the
last, four-spin, diagram is lengthy though its evaluation is not
complicated. We give terms for the hyperfine-mediated and
dipole-dipole interactions separately.

Contributions of the hyperfine-mediated interaction are:

\vskip 0.06 true in \noindent a) the three spin ring with
overlapping corrected,

\begin{eqnarray}
\langle V_{43}^{\rm hf} \rangle = - \sum_{\uparrow \downarrow
\downarrow} B_{ij}^2 B_{jk}^2 \{2it
(\frac{1}{\omega_{ij}^2\omega_{ik}}+\frac{e^{2i\omega_{ij}t}}{\omega_{ij}^2\omega_{jk}})+2
\frac{1-e^{2i \omega_{ij}t}}{\omega_{ij}^3 \omega_{jk}}
+\frac{1-e^{2i \omega_{ik}t}}{\omega_{ik}^2 \omega_{jk}^2} -
\frac{1-e^{2i \omega_{ij}t}}{\omega_{ij}^2 \omega_{jk}^2}\},
 \label{V43hf}
\end{eqnarray}

\vskip 0.06 true in \noindent  b) the diagram compensating a
single site overlapping,

\begin{eqnarray}
\langle V_{4l1}^{\rm hf} \rangle &=& 2 \sum_{\uparrow \downarrow
\downarrow} B_{ij}^2 B_{ik}^2 \{2it
\frac{1+e^{2i\omega_{ij}t}}{\omega_{ij}^2\omega_{ik}}+
\frac{(2\omega_{ik}^2-\omega_{ij}\omega_{ik}+\omega_{ij}^2)(1-e^{2i
\omega_{ij}t})}{\omega_{ij}^3 \omega_{ik}^2\omega_{jk}}\\
\nonumber
 &&-\frac{1-e^{2i(\omega_{ik}+\omega_{ij})t}}{2\omega_{ik}^2 \omega_{ij}^2}
\}+\{j\leftrightarrow k\},
 \label{V4l1hf}
\end{eqnarray}

\vskip 0.06 true in \noindent  c) the compensation of a double
site overlapping (two diagrams),

\begin{eqnarray}
\langle V_{4l2}^{\rm hf} \rangle = - \sum_{\uparrow \downarrow }
B_{ij}^4 \{-2it \frac{1+2e^{2i\omega_{ij}t}}{\omega_{ij}^3}+
\frac{e^{4i\omega_{ij}t}+2e^{2i\omega_{ij}t}-5}{2\omega_{ij}^4}
\}.
 \label{V4l2hf}
\end{eqnarray}

The same diagrams for dipole-dipole interaction only have the
following expressions:

\vskip 0.06 true in \noindent a) the three spin ring with
overlapping corrected,

 \begin{eqnarray}
\langle V_{43}^{\rm dd} \rangle &=& - \sum_{\uparrow \downarrow
\downarrow} B_{ij}^2 B_{jk}^2 \{2it (\frac{2
\omega_{ik}-\omega_{ij}}{\omega_{ij}^2\omega_{ik}\omega_{jk}}-\frac{(1-e^{i\omega_{ij}t})^2}
{\omega_{ij}^2\omega_{jk}})+4\frac{(\omega_{ij}^2-2\omega_{jk}^2)e^{i
\omega_{ij}t}}{\omega_{ij}^3\omega_{ik}\omega_{jk}^2} \\
&&-\frac{e^{2i\omega_{ik}t}}{\omega_{ik}^2\omega_{jk}^2}-
4\frac{e^{i\omega_{ik}t}}{\omega_{ij}\omega_{ik}\omega_{jk}^2}-
\frac{(3\omega_{ij}-2\omega_{jk})e^{2i
\omega_{ij}t}}{\omega_{ij}^3\omega_{jk}^2}+4\frac{e^{i(\omega_{ik}+\omega_{ij})t}}{\omega_{ij}\omega_{ik}\omega_{jk}^2}
+ \frac{\omega_{ij}+6\omega_{ik}}{\omega_{ik}^2\omega_{ij}^3}\},
\nonumber
 \label{V43dd}
\end{eqnarray}

\vskip 0.06 true in \noindent  b) the diagram compensating a
single site overlapping,

\begin{eqnarray}
 \langle V_{4l1}^{\rm dd} \rangle &=& 2 \sum_{\uparrow \downarrow
\downarrow} B_{ij}^2 B_{ik}^2
\{2it\frac{2-(1-e^{i\omega_{ij}t})^2}{\omega_{ij}^2\omega_{ik}}
-\frac{12e^{i(\omega_{ij}+\omega_{ik})t}+e^{2i(\omega_{ij}+\omega_{ik})t}-
8e^{i(\omega_{ik}+2\omega_{ij})}}{2\omega_{ij}^2\omega_{ik}^2}
\\ \nonumber &&
+\frac{8(\omega_{ij}^2+\omega_{ik}^2-\omega_{ij}\omega_{ik})e^{i\omega_{ij}t}
-(2\omega_{ik}^2+3\omega_{ij}^2-3\omega_{ij}\omega_{ik})e^{2i\omega_{ij}t}}{\omega_{ij}^3\omega_{ik}^2(\omega_{ij}-\omega_{ik})}\\
&&
+\frac{6\omega_{ij}^2+\omega_{ij}\omega_{ik}+6\omega_{ik}^2}{2\omega_{ij}^3\omega_{ik}^3}
\} +\{j\leftrightarrow k\} , \nonumber
 \label{V4l1dd}
\end{eqnarray}

\vskip 0.06 true in \noindent  c) the compensation of a double
site overlapping,

\begin{eqnarray}
\langle V_{4l2}^{\rm dd} \rangle = - \sum_{\uparrow \downarrow }
B_{ij}^4 \{2it\frac{2(1-e^{i\omega_{ij}t})^2-3}{\omega_{ij}^3}+
\frac{e^{4i\omega_{ij}t}-8e^{3i\omega_{ij}t}+12e^{2i\omega_{ij}t}+8e^{i\omega_{ij}t}-13}{2\omega_{ij}^4}\}.
 \label{V4l2dd}
\end{eqnarray}

\vskip 0.06 true in \noindent In equations $\{j \leftrightarrow k
\}$ means the same expression with interchanged $j$ and $k$
indexes.

\begin{acknowledgments}
We thank A. M. Tyryshkin and S. A. Lyon for providing us with the
experimental data and for useful discussions. This work was
supported by the National Science Foundation grant DMR-0403465 and
ARO/NSA-LPS.
\end{acknowledgments}


\newpage

\begin{figure}[t]
\centering
\includegraphics[width=3 in, clip=true]{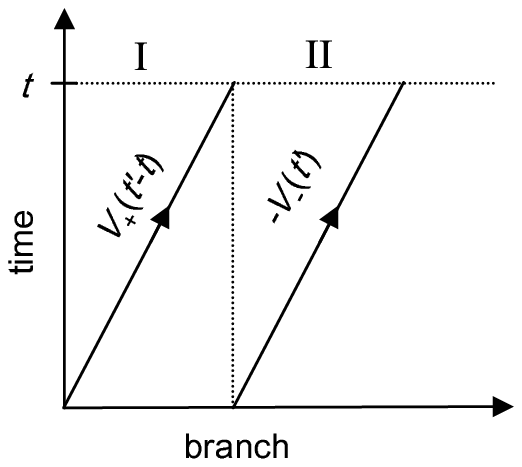}
\caption{} \label{Fig1}
\end{figure}

FIGURE CAPTIONS:

Fig.~\ref{Fig1}: Schematic representation of two-branch evolution.

\begin{figure}[t]
\centering
\includegraphics[width=1.5 in, clip=true]{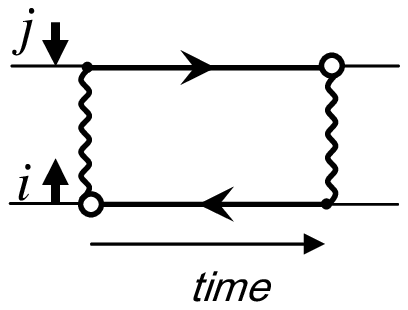}
\caption{} \label{Fig2}
\end{figure}

Fig.~\ref{Fig2}: Second order diagram corresponding to nuclear
spin pair flip-flop processes.

\begin{figure}[t]
\centering
\includegraphics[width=4.5 in, clip=true]{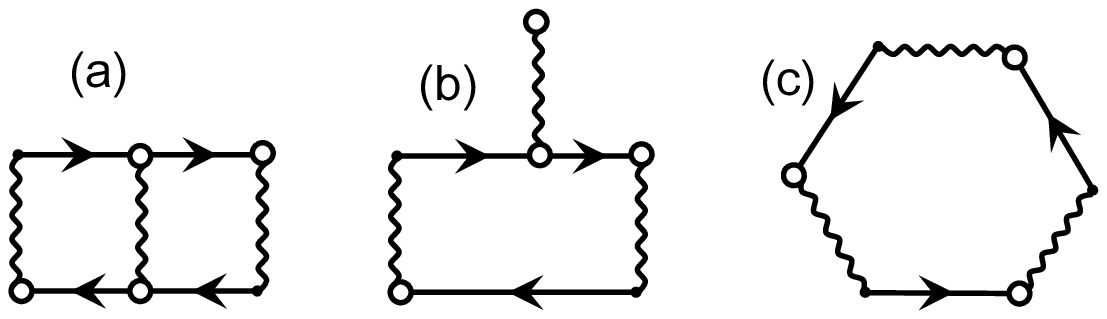}
\caption{} \label{Fig3}
\end{figure}

Fig.~\ref{Fig3}: Set of third order diagrams

\begin{figure}[t]
\centering
\includegraphics[width=4.5 in, clip=true]{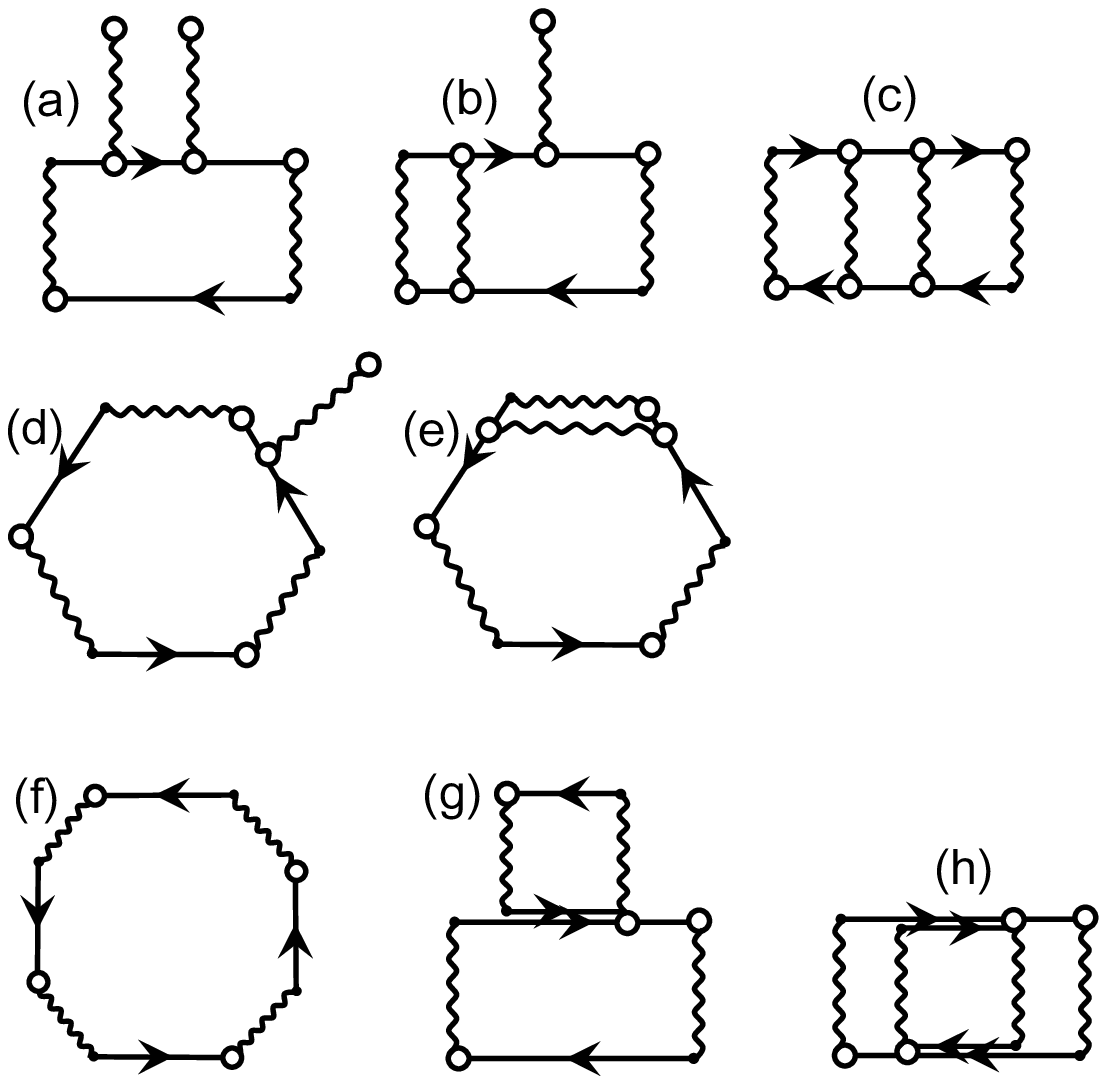}
\caption{} \label{Fig4}
\end{figure}

Fig.~\ref{Fig4}: Fourth order diagrams.

\begin{figure}[t]
\centering
\includegraphics[width=4.5 in, clip=true]{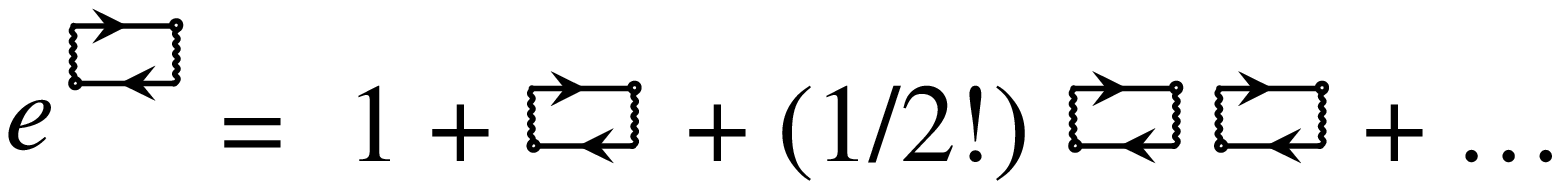}
\caption{} \label{Fig5}
\end{figure}

Fig.~\ref{Fig5}: Second order contribution to decoherence in LCE
that includes a series of even orders of a conventional
perturbation expansion.

\begin{figure}[t]
\centering
\includegraphics[width=4.5 in, clip=true]{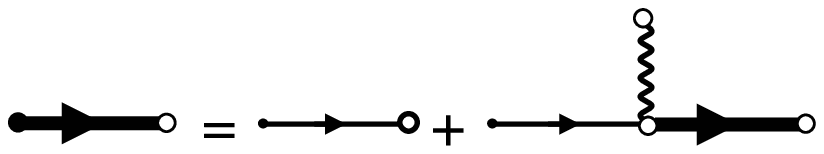}
\caption{} \label{Fig6}
\end{figure}

Fig.~\ref{Fig6}: Renormalization of a Green's function by $I_i^z
I_j^z$ interaction.

\begin{figure}[t]
\centering
\includegraphics[width=5 in, clip=true]{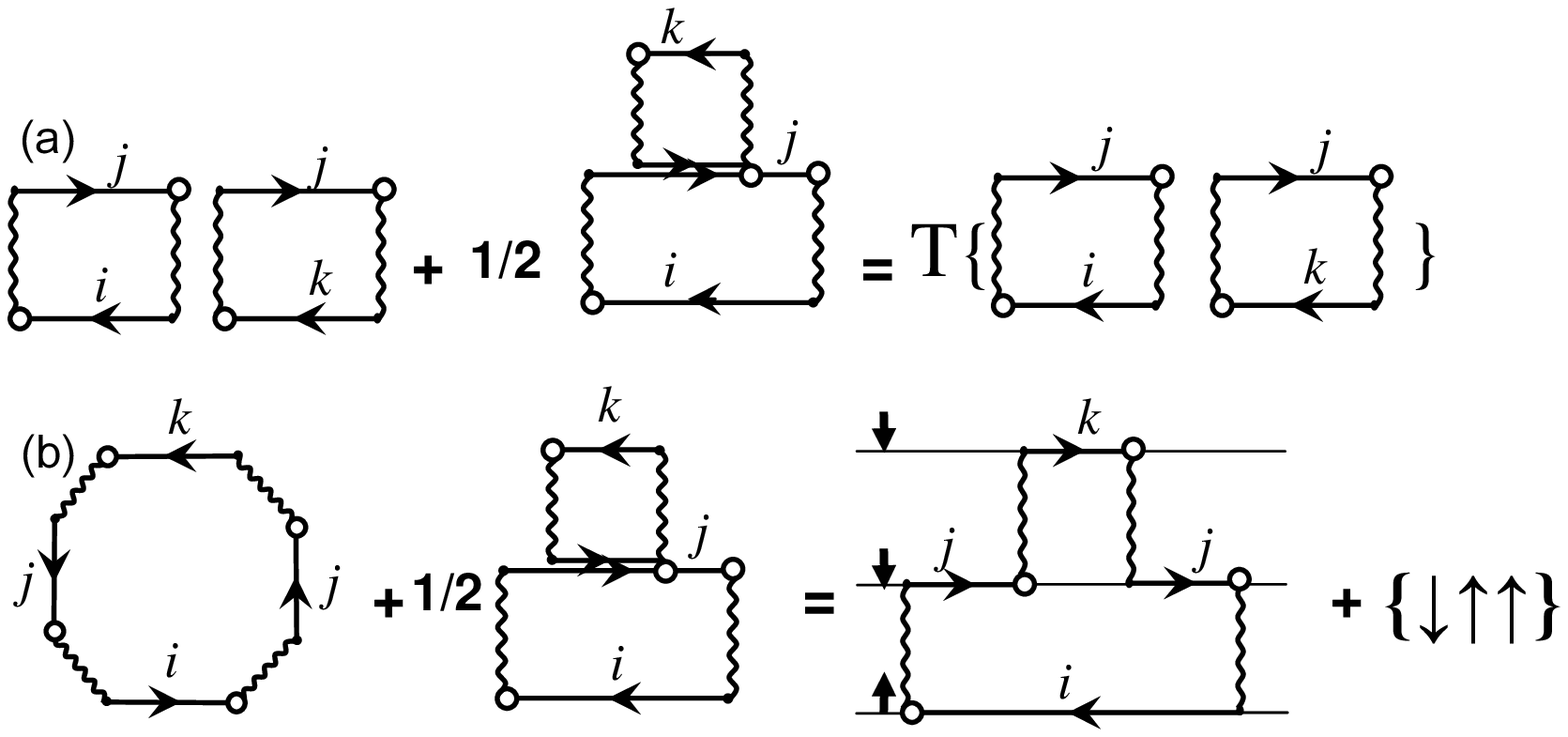}
\caption{} \label{Fig7}
\end{figure}

Fig.~\ref{Fig7}: Compensation of spin excitations overlapping in
time by a locked diagram. Time overlap in one site (j).

\begin{figure}[t]
\centering
\includegraphics[width=5 in, clip=true]{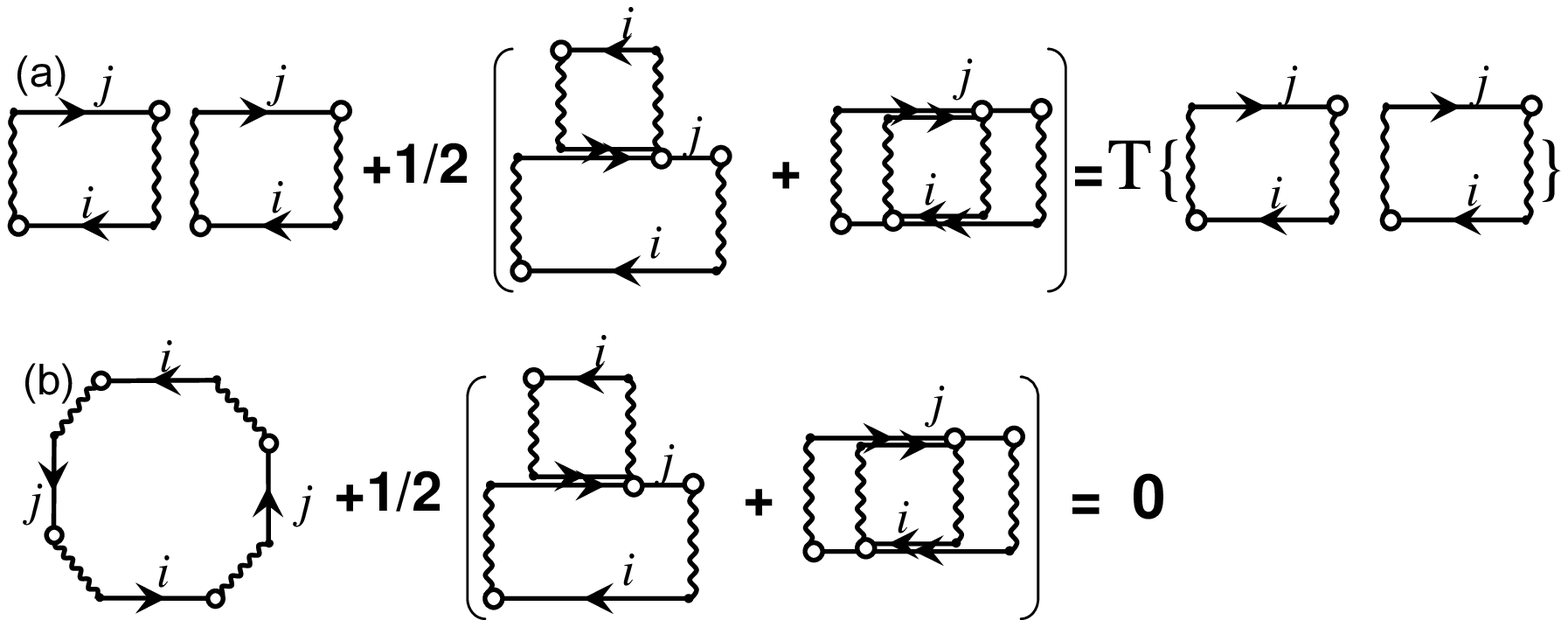}
\caption{} \label{Fig8}
\end{figure}

\begin{figure}[t]
\centering
\includegraphics[width=5 in, clip=true]{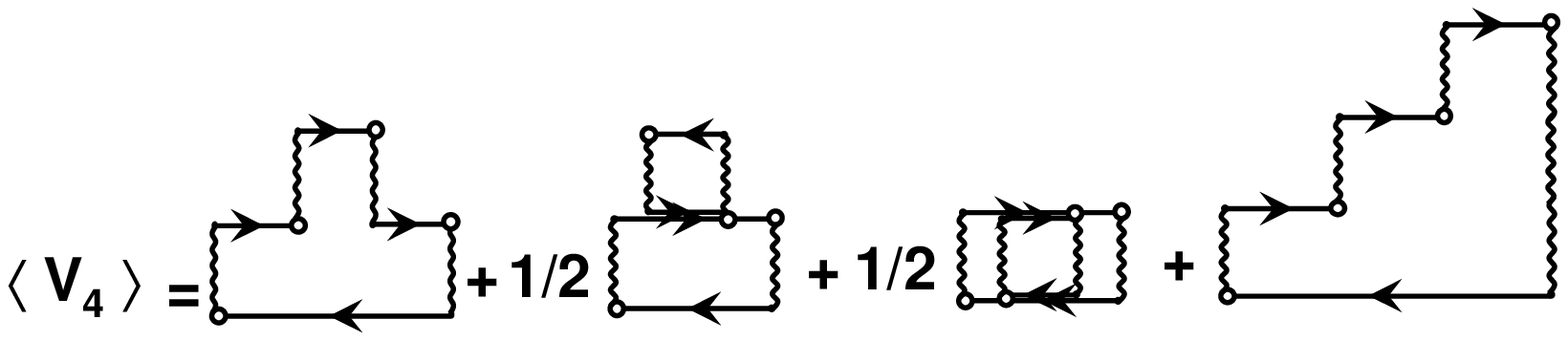}
\caption{} \label{Fig9}
\end{figure}

Fig.~\ref{Fig8}: Compensation of spin excitations overlapping by a
locked diagram. Time overlap in two sites (i and j).

Fig.~\ref{Fig9}: The total fourth order contribution of nuclear
spin excitations to electron spin decoherence. The first diagram
corresponds the ring propagation of spin excitation in three-spin
clusters $\uparrow\downarrow\downarrow$ and
$\downarrow\uparrow\uparrow$. The second, single-site locked,
diagram corrects overlapping of excitations. It contributes spin
dynamics of three and two-spin clusters as shown in
Figs.~\ref{Fig7} and ~\ref{Fig8}. The third, two-site locked,
diagram corrects overlapping of two spin excitations,
Fig.~\ref{Fig8}. The last diagram describes four-spin ring
excitations in spin clusters $\downarrow\uparrow\uparrow\uparrow$,
$\uparrow\downarrow\downarrow\downarrow$,
$\uparrow\downarrow\uparrow\downarrow$ and
$\uparrow\uparrow\downarrow\downarrow$.

\begin{figure}[t]
\centering
\includegraphics[width=4.5 in, clip=true]{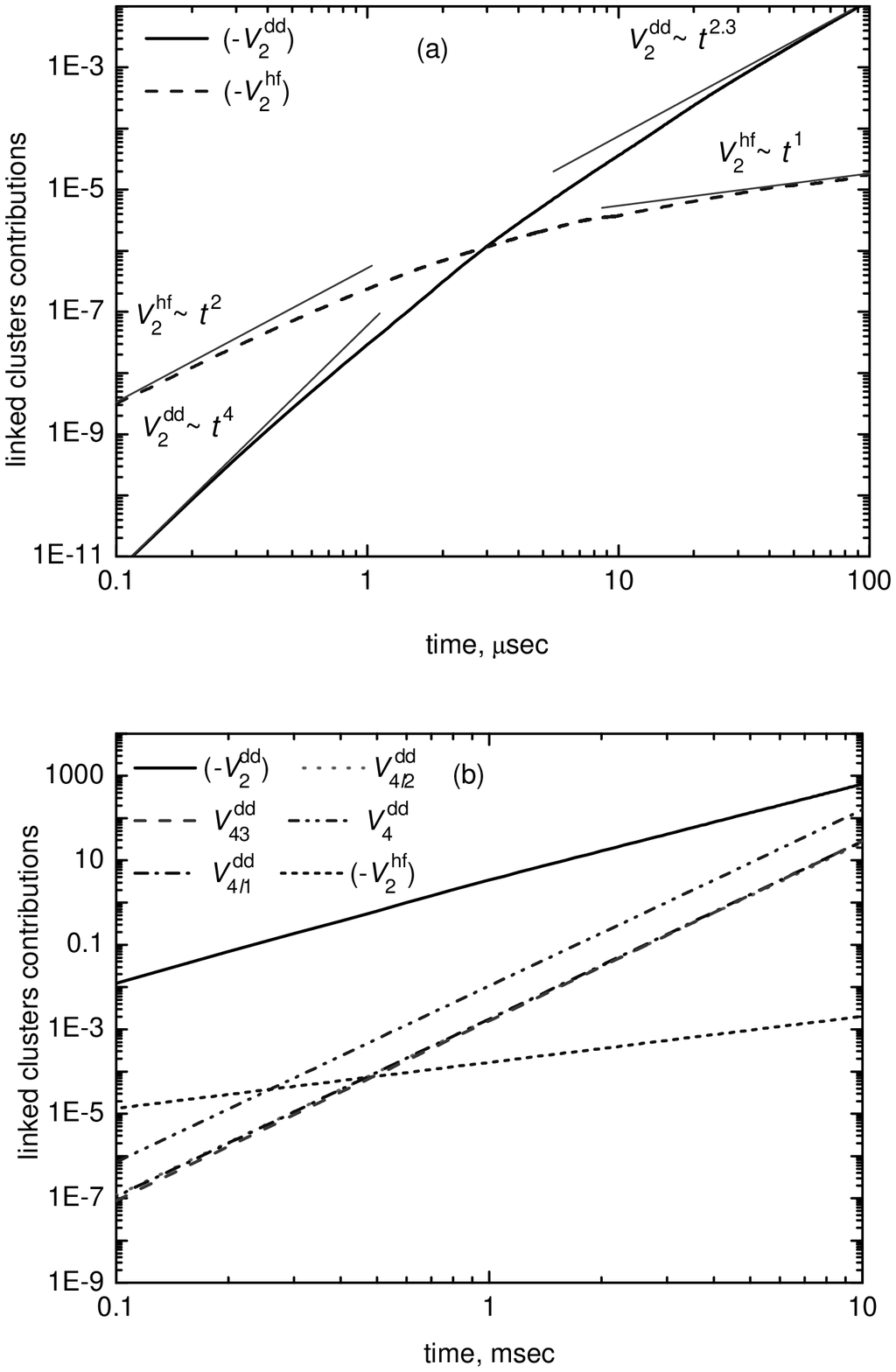}
\caption{} \label{Fig10}
\end{figure}

Fig.~\ref{Fig10}: Contributions of linked clusters to the electron
spin decoherence: (a) Crossover from the short-time evolution to
the intermediate regime. (b) Development of the fourth order terms
in the intermediate time regime. $V_2$ corresponds to the spin
pair flip-flop process in Fig.~\ref{Fig2}; $V_{43}$ - the
three-spin excitation, the first term in Fig.~\ref{Fig9};
$V_{4l1}$ - the single site spin locked diagram, the three-spin
part of the second term in Fig.~\ref{Fig9}; $V_{4l2}$ - correction
of overlapping of two-spin excitations given in the bracket in
Fig.~\ref{Fig8}(a),
and $V_{4}$ - the four-spin ring diagram, the
last term in Fig.~\ref{Fig9}.

\begin{figure}[t]
\centering
\includegraphics[width=4.5 in, clip=true]{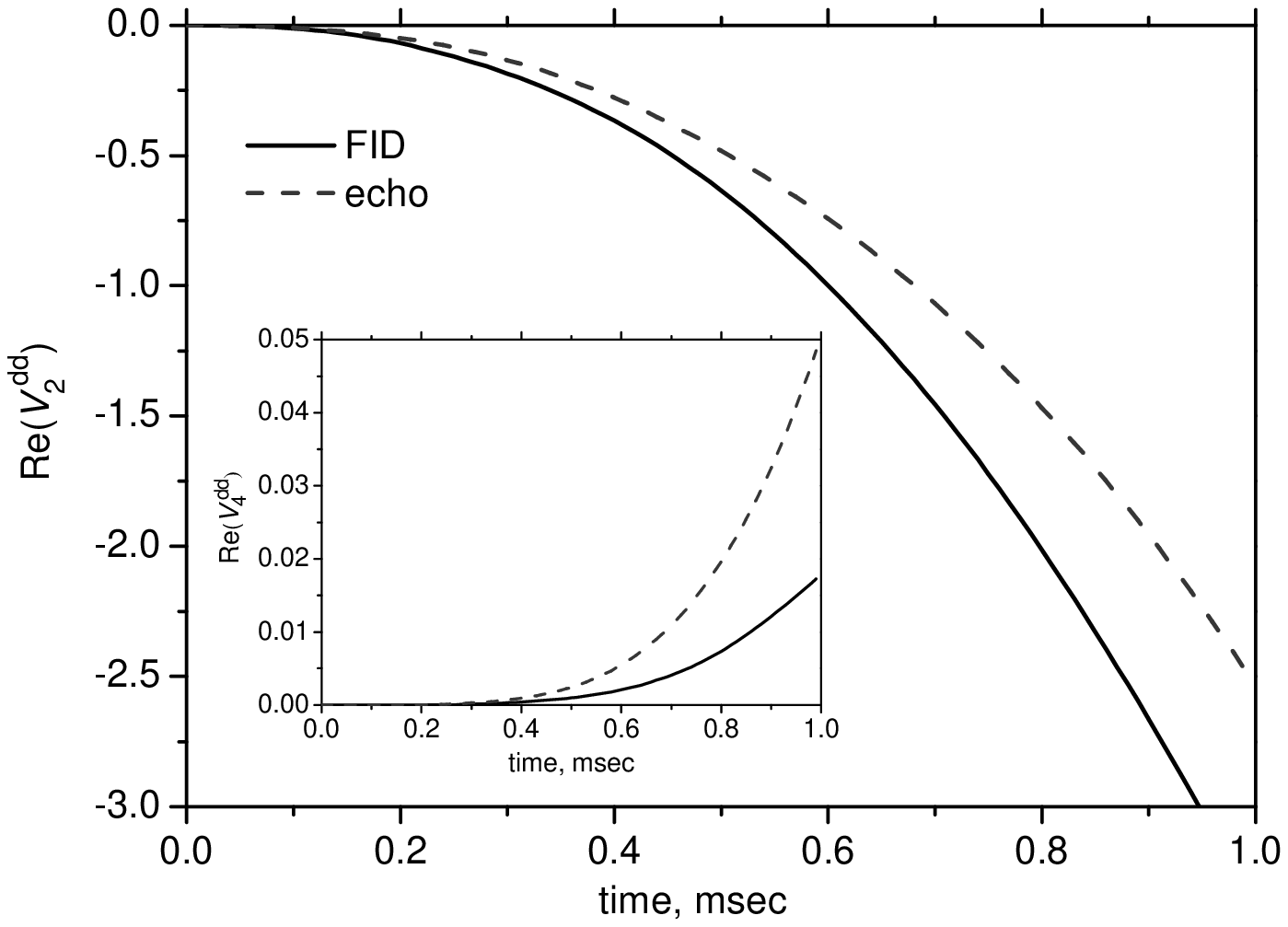}
\caption{} \label{Fig11}
\end{figure}

Fig.~\ref{Fig11}: Time dependence of second and fourth (inset)
order terms in the exponent. FID vs. spin echo decay.

\begin{figure}[t]
\centering
\includegraphics[width=4.5 in, clip=true]{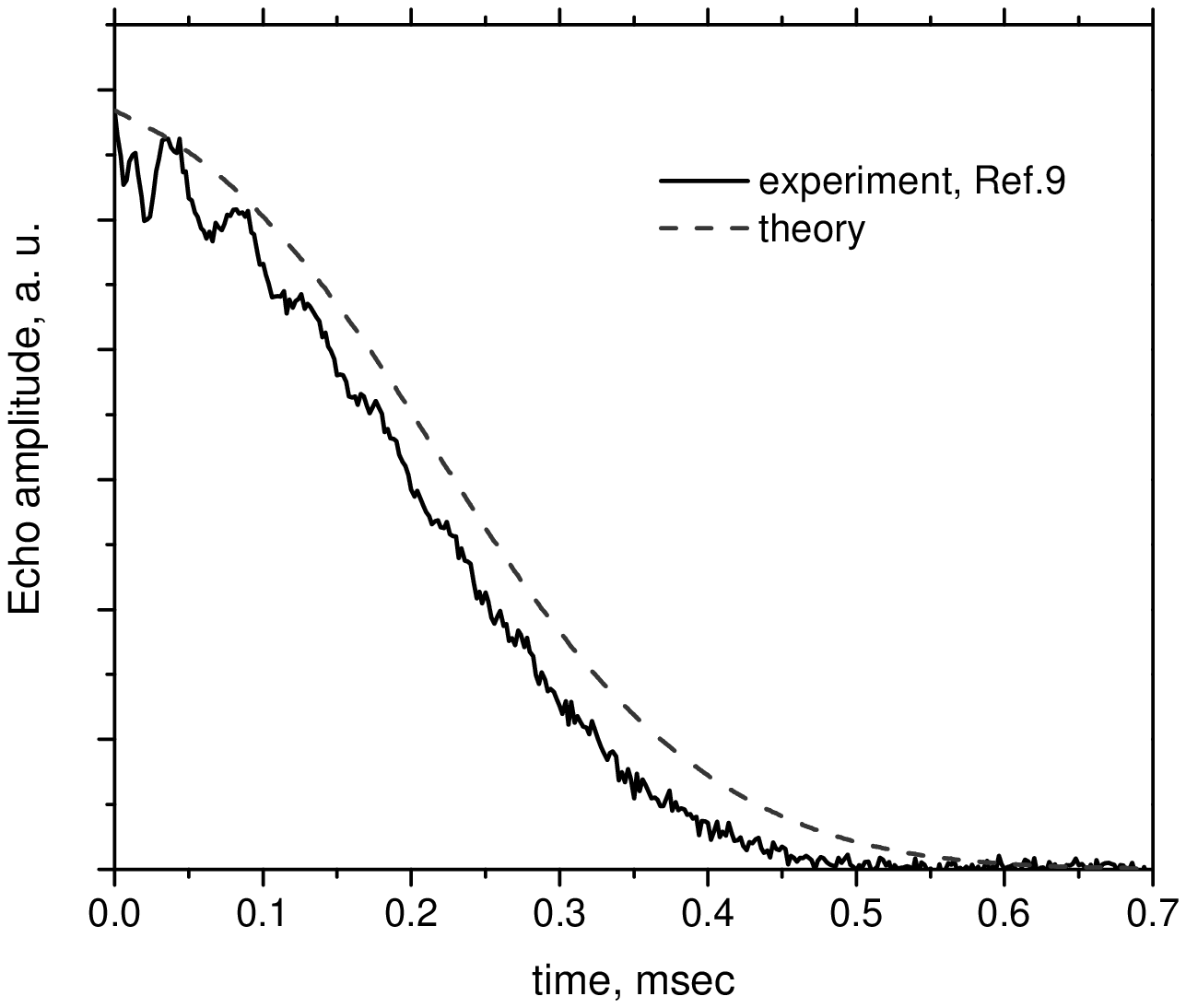}
\caption{} \label{Fig111}
\end{figure}

Fig.~\ref{Fig111}: Spin echo decay in comparison with the
experimental data.\cite{Tyryshk1} Orientation of the external
magnetic field is along (100) direction in the crystallographic
axes.

\begin{figure}[t]
\centering
\includegraphics[width=4.5 in, clip=true]{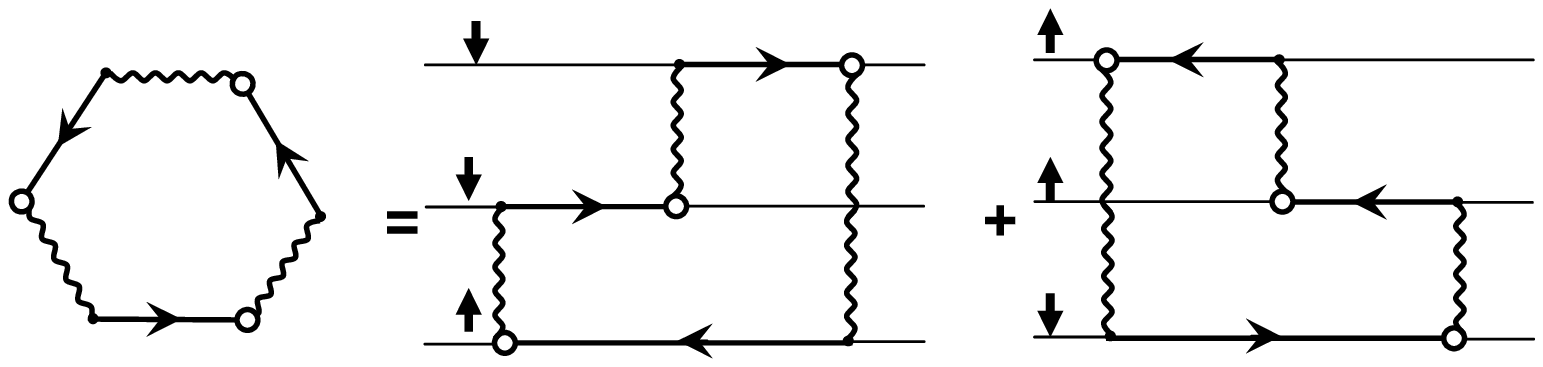}
\caption{} \label{ApA}
\end{figure}

Fig.~\ref{ApA}: Representation of a third order diagram using
cluster configurations.

\newpage
~

\newpage
~

\newpage
~

\newpage
~

\newpage
~

\newpage
~

\newpage
~

\newpage
~

\newpage
~

\newpage
~

\newpage
~

\newpage
~

\newpage
~

\end{document}